\documentclass[12pt,letterpaper]{article}
\usepackage[utf8]{inputenc}
\DeclareUnicodeCharacter{03B8}{\circ}
\usepackage{scalerel}
\usepackage{textcomp}
\usepackage{gensymb}
\usepackage{graphicx}
\usepackage{scalerel}
\usepackage{amsmath}
\usepackage{booktabs}
\usepackage{tabu}
\usepackage{graphicx}
\usepackage{caption}
\usepackage{subcaption}
\usepackage{cite}
\usepackage{secdot}
\usepackage[margin=0.8in]{geometry}
\usepackage{amssymb}
\usepackage{url}
\usepackage[colorlinks = true,
            linkcolor = blue,
            urlcolor  = blue,
            citecolor = blue,
            anchorcolor = blue]{hyperref}\usepackage{ragged2e}
\usepackage{algorithm2e}
\usepackage[affil-it]{authblk}
\usepackage{abstract}

\usepackage[nodisplayskipstretch]{setspace}
\providecommand{\keywords}[1]
{
  \small	
  \textbf{\textit{Keywords---}} #1
}
\begin{document} 

\title{\Large{\textbf{
  Phenomenological Implications of $\boldsymbol{\Delta(54)}$ Flavor Symmetry with Triple Inverse Seesaw
 \vspace{-0.45em}}}}

\author{Hrishi Bora$^1$%
\thanks{\href{mailto:hrishi@tezu.ernet.in}{hrishi@tezu.ernet.in} (Corresponding author)}}
  
   \author{Ng. K. Francis$^{1, 2}$\thanks{\href{mailto:ngkfrancis@mail.jnu.ac.in}{ngkfrancis@mail.jnu.ac.in}}}

\affil{\vspace{-1.05em}$^1$\normalsize{Department of Physics, Tezpur University, Tezpur-784028, India}}
\affil{\vspace{-1.05em}$^2$\normalsize{School of Physical Sciences, Jawaharlal Nehru University, New Delhi-110067, India}}

\date{\vspace{-5ex}}
\maketitle
\begin{center}
\textbf{\large{Abstract}}\\
\justify

We present an extension of the $\Delta(54)$ flavor symmetry model by incorporating two Standard Model Higgs fields. We generated the neutrino mass matrices with the Triple Inverse Seesaw mechanism. Our numerical analysis reveals deviations from tribimaximal neutrino mixing resulting in a nonzero reactor angle ($\theta_{13}$). The atmospheric oscillation parameter ($\theta_{23}$) occupies the upper octant. The model predicts Normal Hierarchy (NH) for neutrino masses. The best-fit value is obtained with minimum $\chi^2$ analysis. The
prediction of the modified model on CP-violating parameters, effective Majorana mass, and Neutrinoless double beta decay phenomenological investigations are found to be in congruous with the current neutrino oscillation data.
\end{center}


\hspace{-1.9em} PACS numbers: 12.60.-i, 14.60.Pq, 14.60.St
\newpage
\section{Introduction}
\label{sec:intro}

The observation of neutrino oscillations exposed a significant shortcoming in the Standard Model, which initially assumed massless neutrinos. Experimental evidence revealed that at least two neutrino flavors possess non-zero masses, validating prior theoretical predictions and highlighting the need for SM extensions.
Three mixing angles are used in the research of neutrino oscillation, two of which are large and one of which is very comparatively small. The reactor mixing angle is not zero, as evidenced by investigations like the Daya Bay Reactor Neutrino Experiment \cite{DayaBay:2012fng} and RENO Experiment\cite{RENO:2012mkc}. The TBM model is unreliable since several other tests, including MINOS\cite{MINOS:2011amj}, Double Chooz\cite{DoubleChooz:2011ymz}, and T2K\cite{T2K:2011ypd}, consistently observed nonzero values for the reactor mixing angle. To accomplish realistic blending, other models or adjustments must be taken into account.

Neutrinoless double beta decay holds the potential to confirm neutrinos as Majorana particles, yet remains undetected. This phenomenon is rooted in Wendell Furry's pioneering work, which introduced the Majorana hypothesis and explored a kinetically analogous process. Both require symmetry principles to explain, laying the groundwork for understanding neutrino properties\cite{Furry:1939qr, DellOro:2016tmg}. This process, expressed as $(A,Z)\rightarrow (A,Z +2)+2e^- $  disrupts the lepton number by two units by generating a pair of electrons, resulting in Majorana neutrino masses. Since neutrino masses are zero in the standard model, the large value of the lepton number violation scale is associated with the smallness of observed neutrino masses($\Lambda \sim 10^{14}-10^{15} $GeV). To generate neutrino masses greater than zero, a model beyond the standard model, such as effective theories employing the Weinberg operator is required. 

Several theoretical frameworks have been developed to achieve tribimaximal mixing (TBM) through the application of non-Abelian discrete symmetries, including $A_4$\cite{barman2023non, barman2023neutrino,Ma:2006km, Vien:2014pta}, $S_3$\cite{Ma:2006km}, $S_4$\cite{thapa2021resonant}, $\Delta(27)$\cite{Ma:2007wu,de2007neutrino,Harrison:2014jqa,CarcamoHernandez:2016piw}, and $\Delta(54)$\cite{loualidi2021trimaximal, ishimori2009lepton, vien2021extension}.To accommodate deviations from TBM, these models incorporate additional flavons. This paper presents a methodology by using the $\Delta(54)$ flavor symmetry framework. The $\Delta$(54) symmetry can manifest itself in heterotic string models on factorizable orbifolds, such as the $T^2/Z_3$ orbifold \cite{nilles2018cp}. In these string models, singlets and triplets are the only representations observed as fundamental modes, with doublets notably absent. However, in magnetized or intersecting D-brane models, doublets can emerge as fundamental modes. We can also suggest an extension to the Standard Model, utilizing $\Delta$(54) symmetry. We have the option to engage with both the singlets ($1_{1}, 1_{2}$) and doublets ($2_{1}, 2_{2}, 2_{3}, 2_{4} $)  representations of $\Delta(54)$, which allow us to represent quarks in different ways. This extension successfully integrates the latest experimental findings on the quark sector, including six quark masses, three mixing angles, and the CP-violating phase\cite{vien2021extension}.

The authors of a prior study proposed the use of the Inverse Seesaw mechanism in conjunction with the $\Delta(54)$ flavor model for Dirac neutrinos\cite{bora2023neutrino}. This study provides a demonstration of the Triple Inverse Seesaw mechanism for Majorana neutrinos, using two Standard Model Higgs bosons. We introduced a Vector-Like (VL) and these additional components produces a Majorana mass term after the symmetry breaking. In order to depart from the prescribed TBM neutrino mixing pattern, we included additional flavons, namely $\chi$, $\chi^\prime$ $\zeta$, $\zeta^\prime$, $\xi$, $\xi^\prime$, $\phi$, $\phi^\prime$ and $\phi^{\prime\prime}$   within the framework of the $\Delta(54)$ symmetry. Furthermore, we have integrated a symmetry of $Z_2 \otimes Z_3 \otimes Z_4$ into our model in order to minimize the presence of undesired components and simplify the process of constructing coupling matrices with certain properties. The structure of the neutrino mass matrix, denoted as $m_{\nu}$, was altered in our study, which pertains to the characterization of neutrino masses. Our investigation primarily focused on symmetry-based analyses. This methodology enables a comprehensive analysis of the neutrino mass ($m_{\nu}$), as well as an exploration of the Jarlskog invariant parameter ($J$) and the neutrinoless double beta decay parameter ($m_{ee}$). The methodology used distinguishes our study from that of other researchers.

The organization of our paper is as follows: The framework of the model, including the fields and their symmetrical transformation characteristics, is outlined in Section \ref{sec: frame}. The neutrino phenomenological findings are numerically analyzed and examined in Section  \ref{num}. In Section  \ref{conc} we
conclude with our final remarks.

\section{Framework of the Model}
\label{sec: frame}
To realize the Majorana triple inverse seesaw mechanism, extending the Standard Model fermion sector is crucial. We achieve this by augmenting the $\Delta(54)$ flavor symmetry model, incorporating two Standard Model Higgs fields ($H$ and $H^\prime$) and distinct flavons. In this study, we have presented two Vector-like (VL) fermions denoted as $N_i$ and $S_i$, which possess the characteristic of being gauge singlets inside the framework of the Standard Model.  The subscript $i$ can be $R$ or $L$ associated with right-handedness and left-handedness respectively.

Our proposed model builds upon the $\Delta(54)$ framework, introducing supplementary flavons to accommodate deviations from the ideal Tri-Bimaximal (TBM) neutrino mixing paradigm\cite{2010}. We put extra symmetry $Z_2\otimes Z_3 \otimes Z_4$ to avoid undesirable terms. Table \ref{tab:2} provides details regarding the composition of the particles and corresponding charge assignment in accordance with the symmetry group. The triplet representation of $\Delta(54)$ is used to assign the left-handed leptons doublets and the right-handed charged lepton. The representations of $\Delta(54)$ symmetry are real that guarantees the construction of the effective Lagrangian.

\begin{table}[ht]
    \centering
    \scalebox{0.8}{
 {\begin{tabular}{c c c c c  c c c c c c c c c c c c c}
    \hline \hline
       \textrm{Field}  &  L & $l $ & $H$ & $H^{\prime}$
  & $N_R$ & $N_L$  & $S_R$ & $S_L$ & $\chi$ & $\chi^\prime$ & $\zeta$ & $\zeta^\prime$ & $\xi$ & $\xi^{\prime}$ & $\phi$ & $\phi^\prime$ & $\phi^{\prime\prime}$\\
     \hline \hline
     \textrm{$\Delta(54)$}  &  $3_{1(1)}$ &  $3_{2(2)} $ & $1_{1}$ & $1_{2}$ & $3_{1(1)}$ & $3_{2(1)}$  & $3_{2(2)}$ & $3_{1(2)}$ & $1_2$ & $2_1$ & $1_{1}$ & $1_{1}$ & $3_{2(1)}$ & $3_{1(1)}$ & $3_{2(2)}$ & $3_{2(1)}$ & $3_{1(2)}$\\
     \textrm{Z}$_2$  &  1 & -1 & 1 & 1 & -1 & 1 & 1 & 1 & -1 & -1 & -1 & 1 & -1 & -1 & 1 & -1 & 1\\
    \textrm{Z}$_3$  &  $\omega$ & $\omega$ & 1 & 1  & 1 & $\omega$  & $\omega$  & $\omega$  & 1 & 1& $\omega$ & 1 & $\omega$ & $\omega$ & 1 & $\omega^2$ & 1\\
\textrm{Z}$_4$  &  1 & -1 & 1 & 1  & 1 & -1  & 1 & 1 & -1 &-1& 1 & -1 & 1 & 1 & 1 &-1& 1\\
\textrm{U(1)}  &  1 & 1 & 0 & 0  & 1 & 1  & 1 & 0 & 0 & 0 & 0 & 0 & 0 & 0& 0 & 0& 0\\
     \hline
    \end{tabular}}}
    \caption{Full particle content of our model}
    \label{tab:1}
    \end{table}

The Lagrangian is as follows \footnote{Considering terms upto dimension-5.}:
\begin{align*}
  \mathcal{L} = & \frac{y_1}{\Lambda} ( l \Bar{L} ) \chi H + \frac{y_2}{\Lambda} ( l \Bar{L} ) \chi^\prime H   + \frac{\Bar{L} \Tilde{H^{\prime}} N_{R}}{\Lambda}y_{\xi} \xi + \frac{\Bar{L} \Tilde{H} N_{R}}{\Lambda} y_{s} \xi^{\prime}     \\                 
 & + y_{\scaleto{NS}{3pt}}  \Bar{S_L} N_R \zeta  + y_{\scaleto{NS}{3pt}}^{\prime} S_{R}\Bar{N_{L}} \zeta^\prime  + y_{\scaleto{S}{3pt}}  S_{L}\Bar{S_{R}} \phi +  y_{\scaleto{N}{3pt}} N_{L}\Bar{N_{R}} \phi^{\prime}   +  \mu S^c_R \Bar{S_R}\phi^{\prime\prime}  + \mu^{\prime} S_L\Bar{S^c_L}\phi^{\prime\prime} + h.c..
\end{align*}

In this context, the vacuum expectation values are considered naturally as,
\begin{align*}
\langle \chi \rangle& =(v_{\chi})&
\langle \chi^\prime \rangle& =(v_{{\chi}^\prime},v_{{\chi}^\prime})&
\langle \xi \rangle& =(v_{\xi}, v_{\xi}, v_{\xi})&
\langle \xi^\prime \rangle& =(v^{\prime}_{\xi} ,v^{\prime}_{\xi},v^{\prime}_{\xi})&
\\
\langle \zeta \rangle& =(v_{\zeta})&
\langle \zeta^\prime \rangle& =(v_{\zeta}^\prime) &  
\langle \phi \rangle& =(v_{\phi},v_{\phi},v_{\phi})&
\langle \phi^\prime \rangle& =(v^\prime_{\phi},v^\prime_{\phi},v^\prime_{\phi}) & \langle \phi^{\prime\prime} \rangle& =(v^{\prime\prime}_{\phi},v^{\prime\prime}_{\phi},v^{\prime\prime}_{\phi})
\end{align*}

The charged lepton mass matrix is given as \cite{ishimori2009lepton} 
\begin{align*}
    M_l= \frac{y_1 v}  {\Lambda}
    \begin{pmatrix}
    v_{\chi} & 0 & 0\\
    0 & v_{\chi}  & 0\\
    0 & 0 &  v_{\chi} 
    \end{pmatrix}  +  
    \frac{y_2 v}  {\Lambda}
    \begin{pmatrix}
    -\omega v_{\chi^\prime} + v_{\chi^\prime} & 0 & 0\\
    0 & -\omega^2 v_{\chi^\prime} + \omega^2 v_{\chi^\prime}  & 0\\
    0 & 0 &   -v_{\chi^\prime} + \omega v_{\chi^\prime}
    \end{pmatrix} 
\end{align*}

where, $y_1$ and $y_2$ are coupling constants and  $v \simeq 55$ GeV.

\subsection{Effective neutrino mass matrix}

The neutrino sector mass matrices are derived from the aforementioned Lagrangian, following the simultaneous breaking of $\Delta(54)$ and electroweak symmetries. Notably, the inverse seesaw mechanism facilitates TeV-scale neutrinos, enabling heavy neutrinos to remain remarkably light ($\sim$TeV) while permitting sizable Dirac masses, comparable to those of charged leptons. This framework reconciles the coexistence of light neutrino masses ($\sim$sub-eV) with appreciable Dirac Yukawa couplings.

 \begin{align}
&M_{NS} =   y_{\scaleto{NS}{3pt}} \begin{pmatrix}
     v_{\zeta} & 0 & 0\\
    0 &  v_{\zeta} & 0\\
    0 & 0 &  v_{\zeta}
    \end{pmatrix};
   & M^{\prime}_{NS}=y^{\prime}_{\scaleto{NS}{3pt}} \begin{pmatrix}
     v^{\prime}_{\zeta} & 0 & 0\\
    0 &  v^{\prime}_{\zeta} & 0\\
    0 & 0 &  v^{\prime}_{\zeta}
    \end{pmatrix}   \\
&M_{S}=y_{\scaleto{S}{3pt}} \begin{pmatrix}
     v_{\phi} & 0 & 0\\
    0 & v_{\phi}  & 0\\
    0 & 0 &  v_{\phi} 
    \end{pmatrix};
&M_{N}=   y_{\scaleto{N}{3pt}} \begin{pmatrix}
     v^{\prime}_{\phi} & 0 & 0\\
    0 & v^{\prime}_{\phi}  & 0\\
    0 & 0 &  v^{\prime}_{\phi} 
    \end{pmatrix}    \\
&M_\mu=  y_\mu\begin{pmatrix}
     v^{\prime\prime}_{\phi} & 0 & 0\\
    0 &  v^{\prime\prime}_{\phi} & 0\\
    0 & 0 &   v^{\prime\prime}_{\phi}
    \end{pmatrix} ;
 &M_\mu^{\prime}=  y^{\prime}_\mu
 \begin{pmatrix}
    v^{\prime\prime}_{\phi} & 0 & 0\\
    0 & v^{\prime\prime}_{\phi} & 0\\
    0 & 0 &  v^{\prime\prime}_{\phi}
    \end{pmatrix}
    \end{align}
 \begin{equation}
 M_{\nu N}= \frac{v}{\Lambda}        \begin{pmatrix} 
    y_{\scaleto{\xi}{6pt}} v_{\scaleto{\xi}{6pt}} & y_{s}v^{\prime}_{\xi}  & y_{s}v^{\prime}_{\xi} \\
   y_{s}v^{\prime}_{\xi} &   y_{\scaleto{\xi}{6pt}} v_{\scaleto{\xi}{6pt}} & y_{s}v^{\prime}_{\xi} \\
    y_{s}v^{\prime}_{\xi}  & y_{s}v^{\prime}_{\xi} &   y_{\scaleto{\xi}{6pt}} v_{\scaleto{\xi}{6pt}}
    \end{pmatrix}
   \end{equation}\\
    The model parameters
naturally follow the already familiar inverse seesaw hierarchy\\
\begin{equation}
  M_{NS}, M_{\mu}, M^{\prime}_\mu, M^{\prime}_{NS}
 << M_{\nu N} << M_N,M_S  
\end{equation}
Using this hierarchy we obtain the effective neutrino mass matrix as \\  
\begin{equation} 
m_\nu = M^2_{\nu N} \frac{M^{\prime 2}_{NS} M^{\prime}_{\mu}}{M^2_{N}M^2_{S}-2M_{S}M_{N} M^{\prime}_{NS} M_{NS} +  M^{\prime 2}_{NS} M^2_{NS} - M^2_{N} M^2_{\mu} M^{\prime 
2}_{\mu}} 
\approx 
M^2_{\nu N}\frac{M^{\prime 2}_{NS} M^{\prime}_\mu}{M^2_{N}M^2_{S}}
\end{equation}

\begin{equation}
  m_\nu=  \lambda
    \begin{pmatrix}
     2a^2 + b^2   &   a^2 + 2ab   &      a^2 + 2ab\\
     a^2 + 2ab    &  2a^2 + b^2   &       a^2 + 2ab \\
    a^2 + 2ab &   a^2 + 2ab   &    2a^2 + b^2 
    \end{pmatrix}\\   
\end{equation}
where $\lambda = \frac{v^2 v^{\prime 2}_{\zeta} y^{\prime 2}_{\scaleto{NS}{3pt}} y^{\prime}_{\mu}v^{\prime\prime}_\phi} {  \Lambda^2 v^2_{\phi} v^{\prime2}_{\phi} y^{2}_{\scaleto{N}{3pt}} y^{\prime 2}_{\scaleto{S}{3pt}}} $, $a=y_{\xi} v_{\xi}$ and $b=y_{\scaleto{\xi}{6pt}} v_{\scaleto{\xi}{6pt}}$.
 
In direct analogy with the standard inverse seesaw, contributions coming from $M_{NS}$ and $M_\mu$ can
be safely neglected. Note that here the suppression mechanism is enhanced by three matrices, $M^{\prime 2}_{NS}$ and $M^{\prime}_{\mu}$.
Therefore, we would call this mechanism triple Majorana inverse seesaw \cite{chulia2021inverse}.

\section{Numerical Analysis and results} 
\label{num}
This section provides a quantitative analysis of the efficacy of flavon-driven corrections to Tri-Bimaximal (TBM) mixing within the $\Delta(54)$ framework, with a focus on the normal hierarchical neutrino mass ordering.

\noindent The neutrino mass matrix $m_\nu$ can be diagonalized by the PMNS matrix $U$ as
\begin{equation}
    \label{eq:12}
    U^\dagger m_\nu U^* = \textrm{diag(}m_1, m_2, m_3 \textrm{)}
\end{equation}
 We can numerically calculate $U$ using the relation $U^\dagger M_\nu U = \textrm{diag(}m_1^2, m_2^2, m_3^2 \textrm{)}$, where $M_\nu = m_\nu m^{\dagger}_\nu$. The neutrino oscillation parameters $\theta_{12}$, $\theta_{13}$, $\theta_{23}$ and $\delta$ can be obtained from $U$ as
\begin{equation}
    \label{eq:13}
    s_{12}^2 = \frac{\lvert U_{12}\rvert ^2}{1 - \lvert U_{13}\rvert ^2}, ~~~~~~ s_{13}^2 = \lvert U_{13}\rvert ^2, ~~~~~~ s_{23}^2 = \frac{\lvert U_{23}\rvert ^2}{1 - \lvert U_{13}\rvert ^2}
\end{equation}

and $\delta$ may be given by
\begin{equation}
    \label{eq:14}
    \delta = \textrm{sin}^{-1}\left(\frac{8 \, \textrm{Im(}h_{12}h_{23}h_{31}\textrm{)}}{P}\right)
\end{equation}
with 
\begin{equation}
    \label{eq:15}
     P = (m_2^2-m_1^2)(m_3^2-m_2^2)(m_3^2-m_1^2)\sin 2\theta_{12} \sin 2\theta_{23} \sin 2\theta_{13} \cos \theta_{13}
\end{equation}

\begin{table}[ht]
\centering
 \scalebox{1}{
  \begin{tabular}{ l  c  r }
    \hline
    Parameters & NH (3$\sigma$) & IH (3$\sigma$) \\  \hline
    $\Delta{m}^{2}_{21}[{10}^{-5}eV^{2}]$ & $6.92 \rightarrow 8.05$ & $6.92 \rightarrow 8.05$ \\ 
    $\Delta{m}^{2}_{31}[{10}^{-3}eV^{2}]$ & $2.463 \rightarrow 2.606$ & $-2.584 \rightarrow -2.438 $\\ 
    $\sin^{2}\theta_{12}$ & $0.275 \rightarrow 0.345$ & $0.275 \rightarrow 0.345$ \\ 
     $\sin^{2}\theta_{13}$ & $0.02023 \rightarrow 0.02376$ & $0.02053 \rightarrow 0.02397$ \\ 
    $\sin^{2}\theta_{23}$ & $0.430 \rightarrow 0.596$ & $0.437 \rightarrow 0.597$ \\
    $\delta^{\circ}_{CP}$ & $96 \rightarrow 422$ & $201 \rightarrow 348$ \\ \hline
  \end{tabular}}
  \caption{ The neutrino oscillation parameters from NuFIT 6.0 (2024) \cite{estebannufit}}
    \label{tab:2}
\end{table}

The 3$\sigma$ ranges of neutrino oscillation parameters from NuFIT 6.0 \cite{estebannufit}. We adjusted the modified $\Delta(54)$ model to suit the experimental data by minimizing the ensuing $\chi^2$ function in order to evaluate how the neutrino mixing parameters contrast with the most current experimental data:

\begin{equation}
	\label{eq:16}
	\chi^2 = \sum_{i}\left(\frac{\lambda_i^{model} - \lambda_i^{expt}}{\Delta \lambda_i}\right)^2,
\end{equation}

where $\lambda_i^{model}$ is the $i^{th}$ observable predicted by the model, $\lambda_i^{expt}$ stands for  $i^{th}$ experimental best-fit value and $\Delta \lambda_i$ is the 1$\sigma$ range of the observable.

 The best-fit values for $\lvert b\rvert$, $\lvert a\rvert$, $\phi_b$ and $\phi_a$  obtained are (0.0124, 
1.6354, -1.2841$\pi$, 0.5123$\pi$).

\begin{figure}[t]
     \centering
     \begin{subfigure}[b]{0.54\textwidth}
         \centering
         \includegraphics[width=\textwidth]{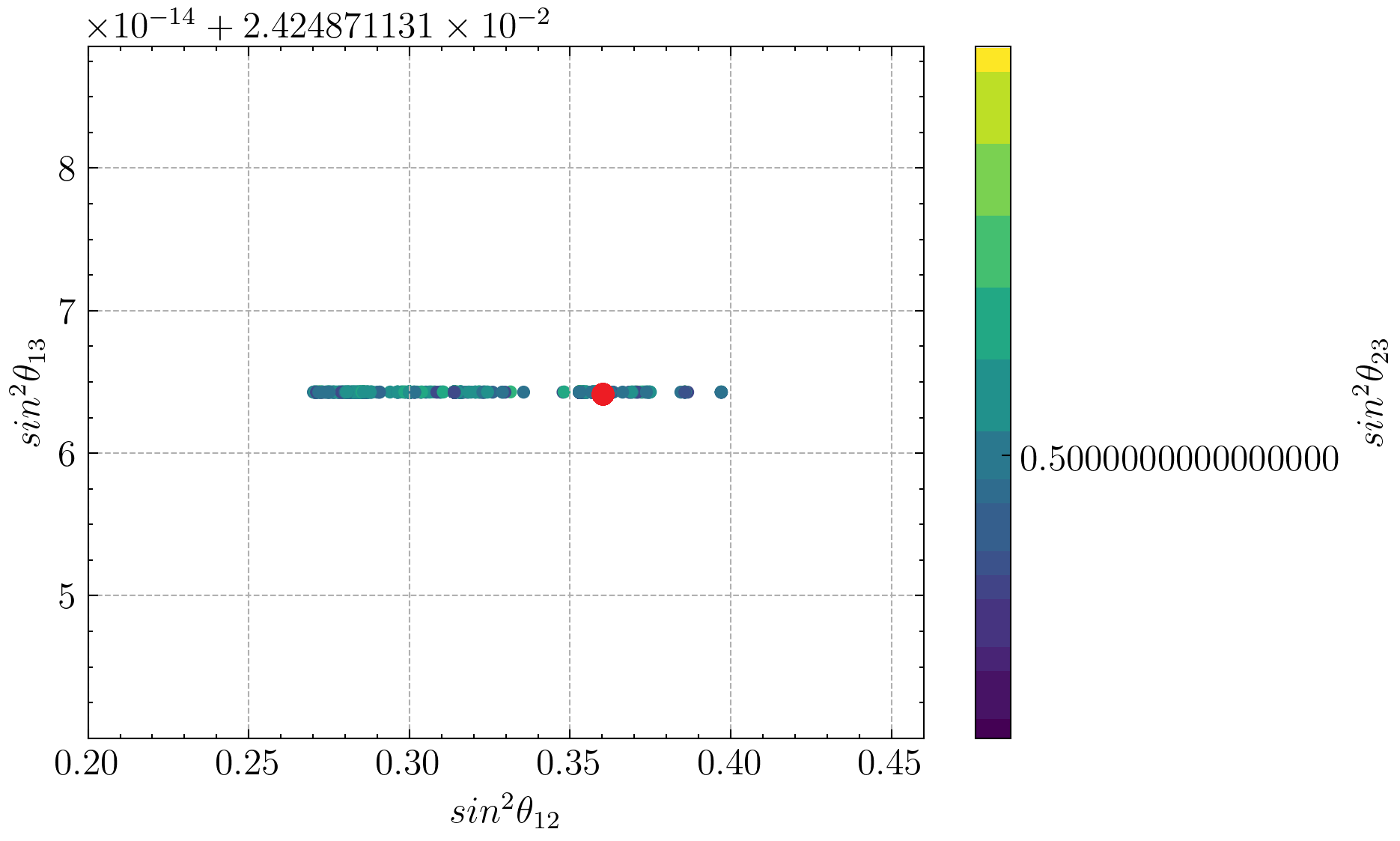}
     \end{subfigure}
     \hfill
     \begin{subfigure}[b]{0.42\textwidth}
         \centering
         \includegraphics[width=\textwidth]{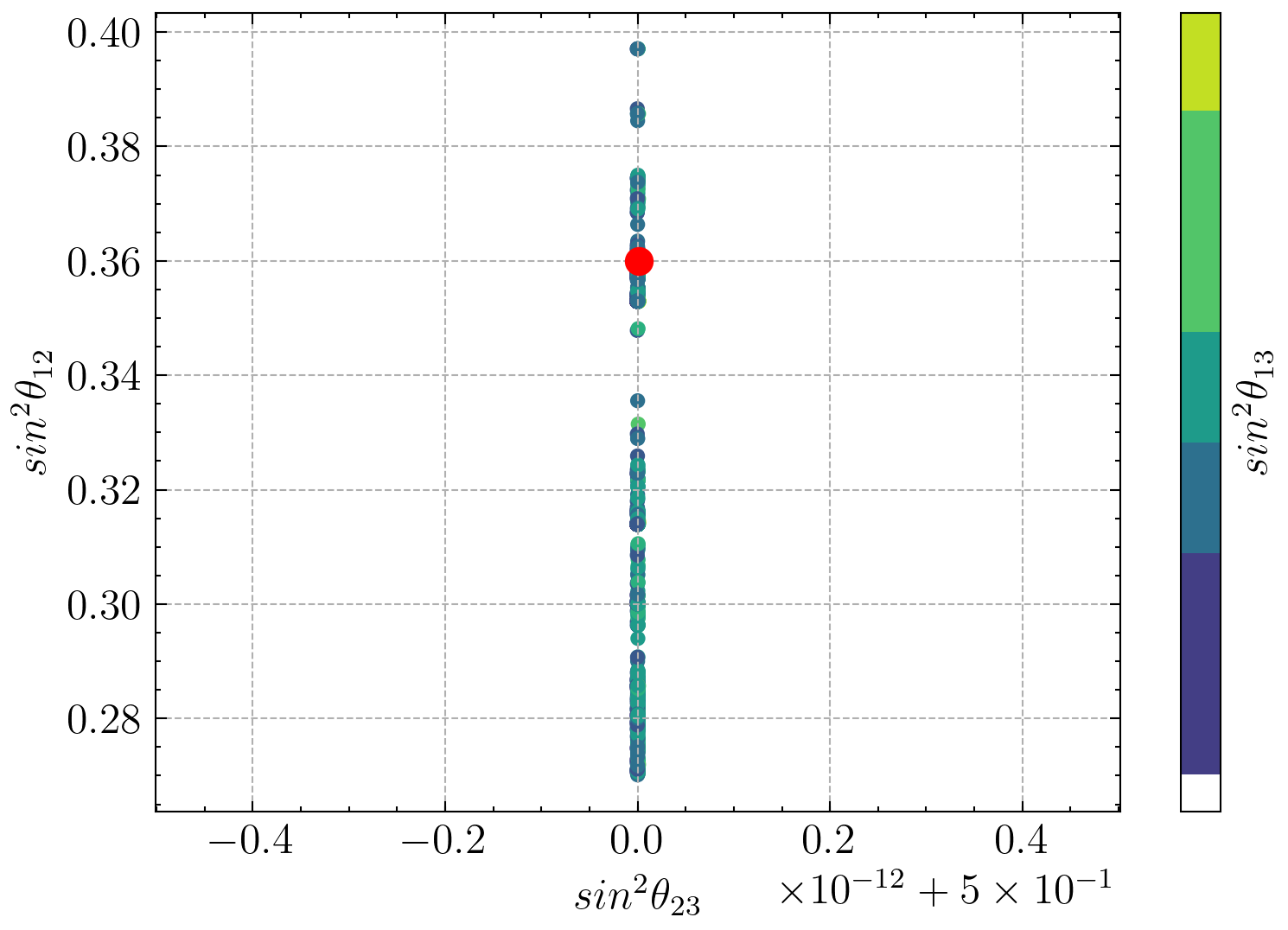}
     \end{subfigure}

     \vspace{1em}
     \begin{subfigure}[b]{0.42\textwidth}
         \centering
         \includegraphics[width=\textwidth]{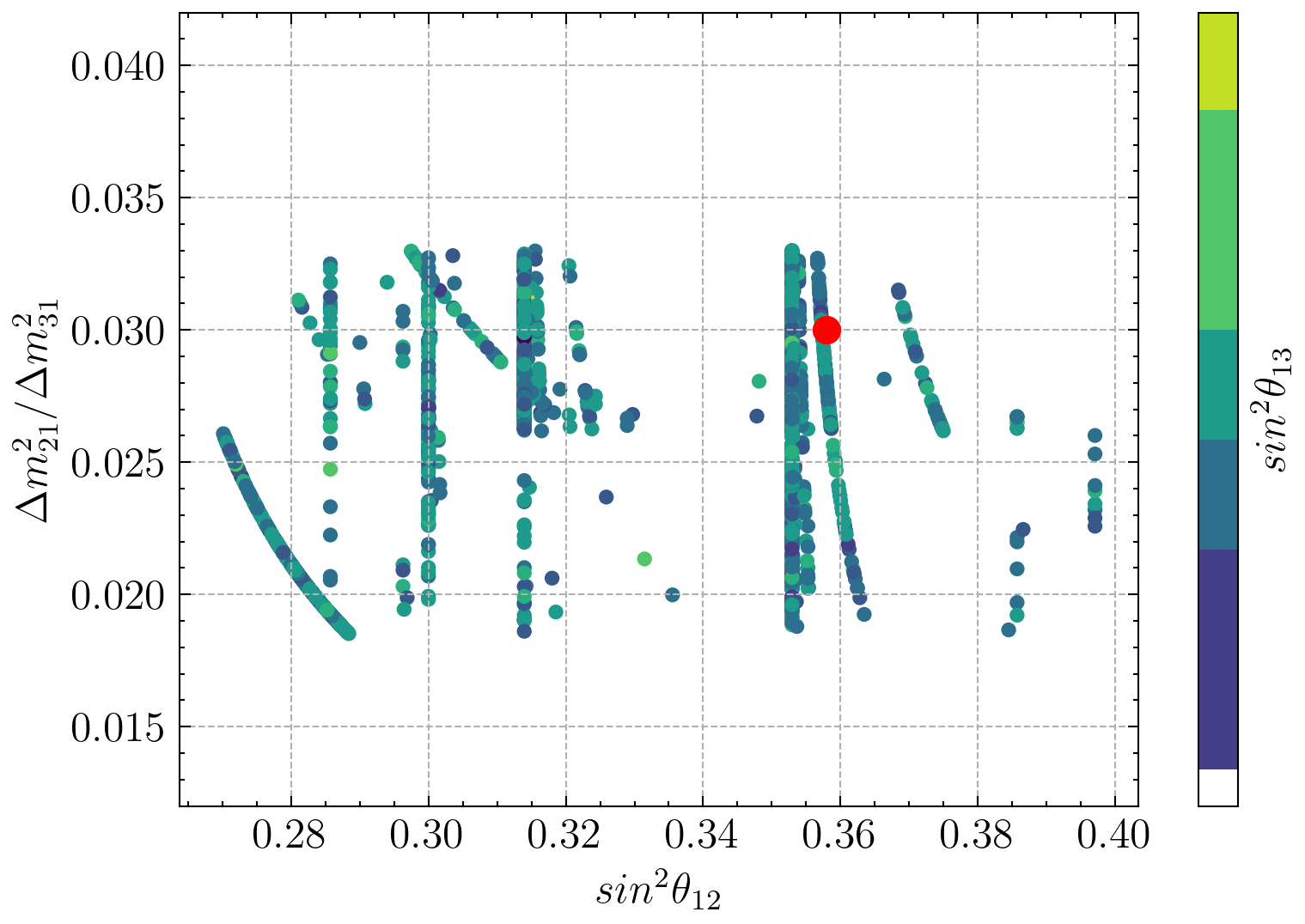}
     \end{subfigure}
    \caption{Correlation among the oscillation parameters predicted by the model at $3\sigma$ in normal hierarchy. The best fit value is indicated by the red dot.}
    \label{fig:2}
\end{figure}

The anticipated values of the neutrino oscillation parameters for NH are shown in Fig.\ref{fig:2}. The values of $sin^2\theta_{12}$, $sin^2 \theta_{13}$, and $sin^2 \theta_{23}$ that best suit the experimental measurements  3$\sigma$ range are 0.36, 0.024, and 0.500002 respectively. The atmospheric mixing angle $sin^2\theta_{23}$ is found to be slightly greater than 0.5 indicating that it resides in the upper octant. The best-fit values for other parameters, such as  $\Delta m^2_{21}/\Delta m^2_{31}$ , are correspond to the $\chi^2$-minimum.

\begin{figure}[t]
     \centering
     \begin{subfigure}[b]{0.4\textwidth}
         \centering
         \includegraphics[width=\textwidth]{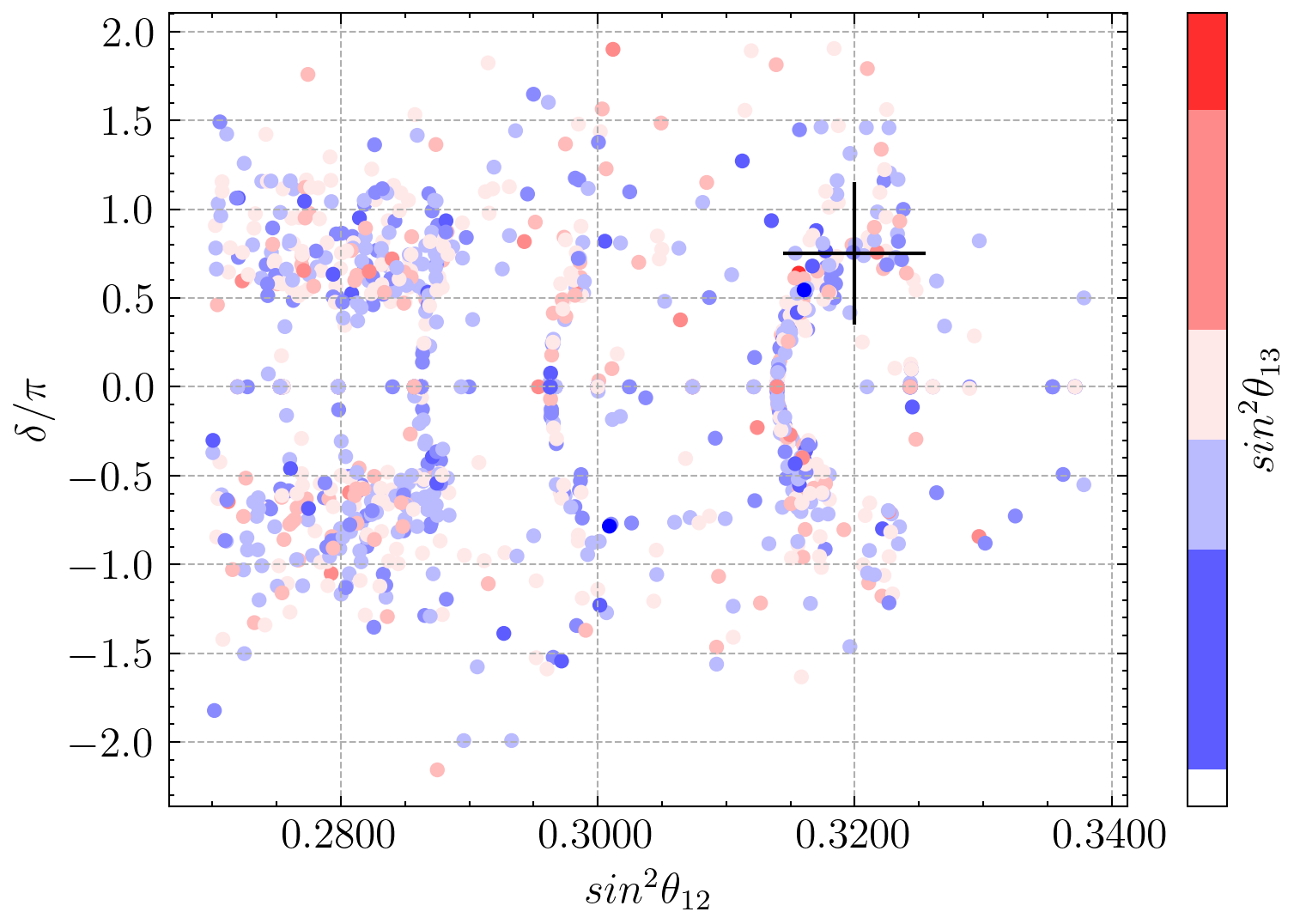}
     \end{subfigure}
     \hfill
     \begin{subfigure}[b]{0.4\textwidth}
         \centering
         \includegraphics[width=\textwidth]{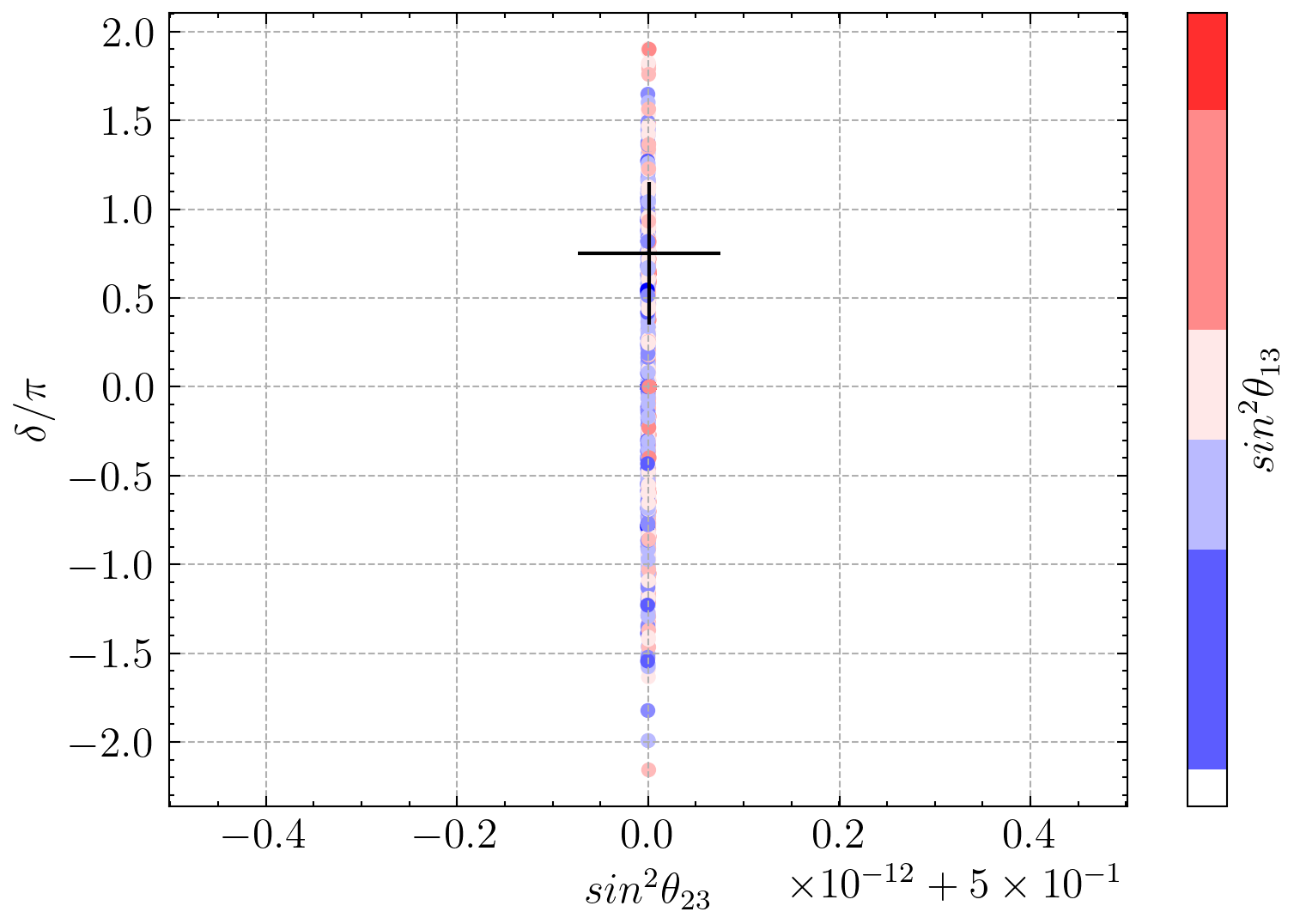}
     \end{subfigure}

    \caption{Correlation between Dirac CP ($\delta_{CP}$) with solar mixing angle($sin^2\theta_{12}$) and atmospheric mixing angle($sin^2\theta_{23}$) respectively. The best fit value for 3$\sigma$ range is given by the + marker.}
    \label{fig:3}
\end{figure}

Fig. \ref{fig:3} gives the correlation between the CP phase with reactor mixing angle and atmospheric  mixing angle respectively. Thus, the model defined in this work indicates clear deviation from tri-bimaximal mixing. The best fit value of $\delta_{CP}$ is predicted to be around $0.75 \pi$.

The correlation between the neutrino oscillation parameters makes it obvious that the neutrino mixing differs from the TBM mixing in NH. The upper octant is preferred in the NH scenario, according to the prediction of mixing angle $\theta_{23}$. It is feasible to depart from TBM mixing by modifying the $\Delta(54)$ model.

\begin{figure}[!ht]
     \centering
     \begin{subfigure}[b]{0.4\textwidth}
         \centering
         \includegraphics[width=\textwidth]{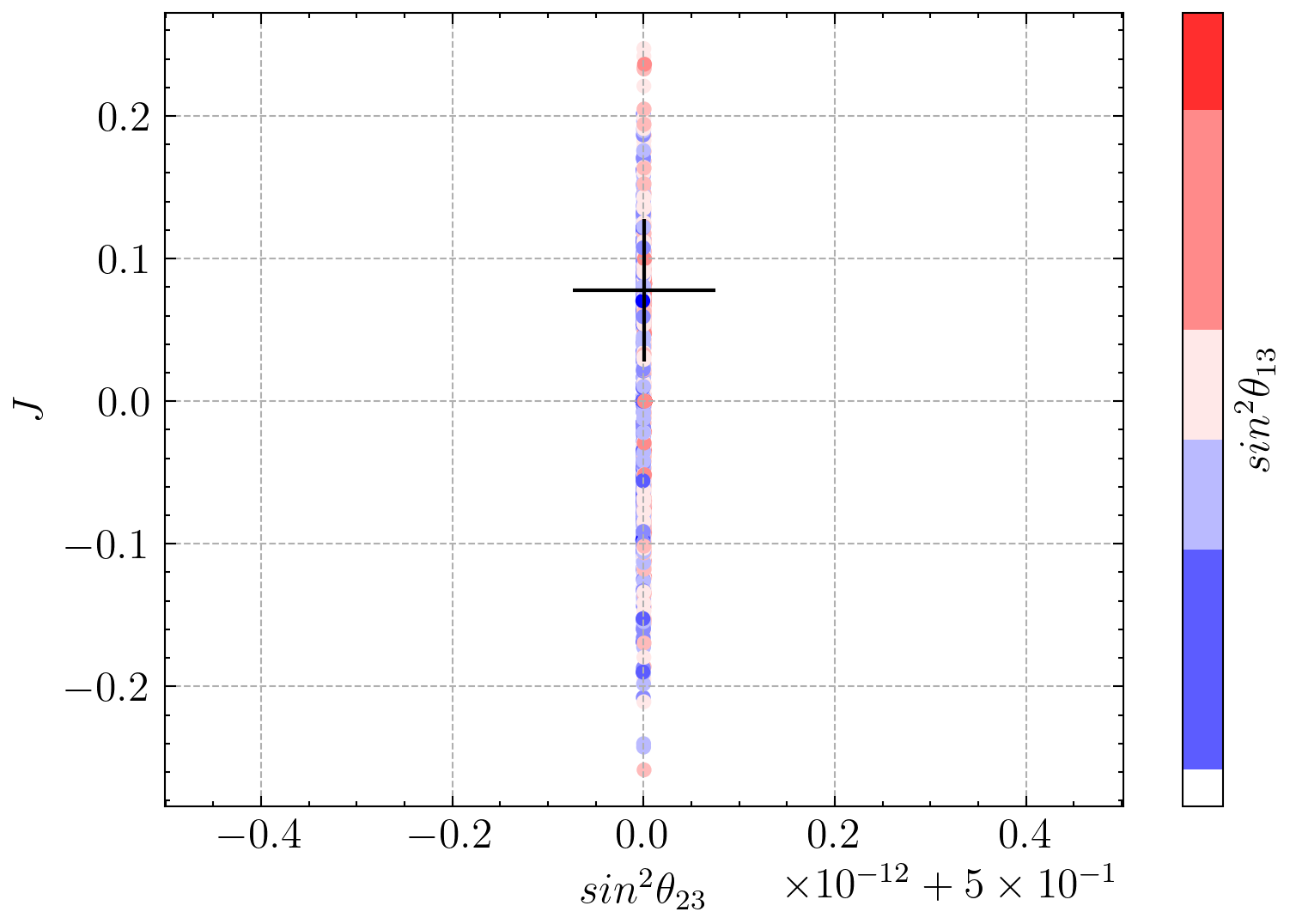}
     \end{subfigure}
     \hfill
     \begin{subfigure}[b]{0.4\textwidth}
         \centering
         \includegraphics[width=\textwidth]{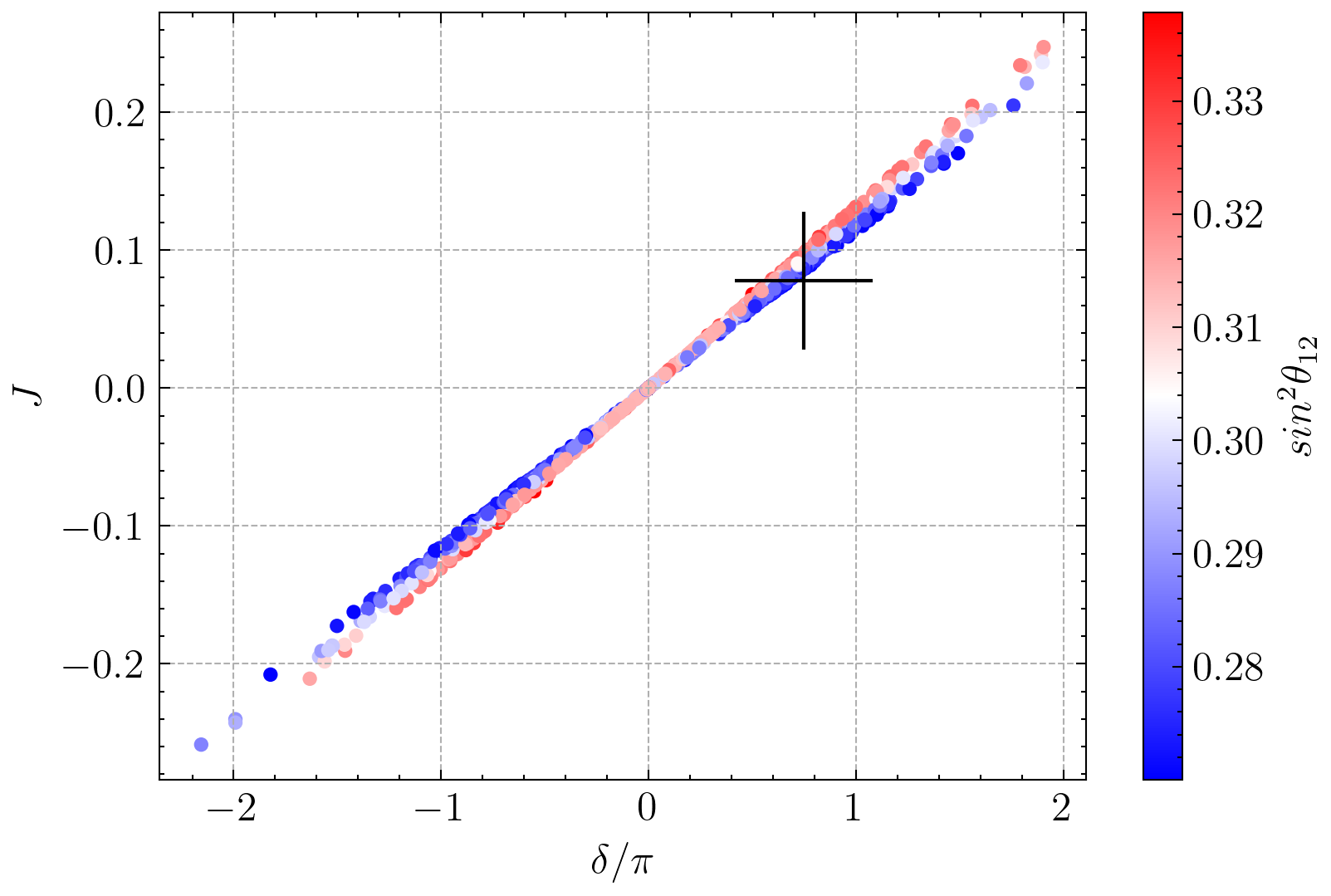}
     \end{subfigure}
    \caption{Correlation between the Jarlskog invariant parameter($J$) with atmospheric mixing angle($sin^2\theta_{23}$) and CP phase ($\delta_{CP}$) and respectively. The best fit value for 3$\sigma$ range is given by the + marker.}
    \label{fig:4}
\end{figure}

\noindent \textbf {Jarlskog invariant Parameter:}
In Fig. \ref{fig:4} we further estimate the  CP-violation with Jarlskog parameter ($J$). This parameter is completely defined by the mixing angles and the Dirac phase. The Jarlskog constant is a quantity that remains unchanged even after a phase redefinition\cite{lei2020minimally}.
\begin{equation}
J= Im \{{U_{11}U_{22} U_{12}^{*} U_{21}^{*}}\} = s_{12}c_{13}^{2}s_{12}c_{12}s_{23}c_{23} sin\delta 
\end{equation}

\noindent \textbf {Neutrinoless double beta decay (NDBD):}
The Neutrinoless Double Beta Decay (NDBD) phenomenon holds profound implications for neutrino physics, as it is intimately linked to the properties of light Majorana neutrinos. This process is governed by the effective Majorana mass $\lvert m_{ee} \rvert$, which can be calculated using the following equation:

\begin{equation}
\lvert m_{ee} \rvert = U^2_{Li} m_i
\end{equation}
 where $U_{Li}$ are the elements of the first row of the neutrino mixing matrix $U_{PMNS}$. This equation depends on certain known parameters such as $\theta_{12}$ and $\theta_{13}$, as well as unknown Majorana phases denoted by $\alpha$ and $\beta$. The diagonalizing matrix of the light neutrino mass matrix, denoted by $m_\nu$, is represented by $U_{PMNS}$, such that 
 \begin{equation}
  m_\nu= U_{PMNS} M^{(diag)}_\nu U^T_{PMNS}
 \end{equation}
  where, $M^{(diag)}_\nu$ =diag($m_1$, $m_2$, $m_3$). The effective Majorana mass can be expressed applying the diagonalizing matrix elements and the mass eigenvalues as follows:
  \begin{equation}
   \lvert  m_{ee} \rvert = m_1 c_{12}^2 c^2_{13}+ m_2 s^2_{12} c^2_{13} e^{2\iota \alpha} + m_3 s^2_{13} e^{2\iota \beta}
  \end{equation} where $c_{12}$ and $s_{12}$ are the cosine and sine of the mixing angle $\theta_{12}$, respectively.

\begin{figure}[!ht]
     \centering
     \begin{subfigure}[b]{0.46\textwidth}
         \centering
         \includegraphics[width=\textwidth,  height = 6cm]{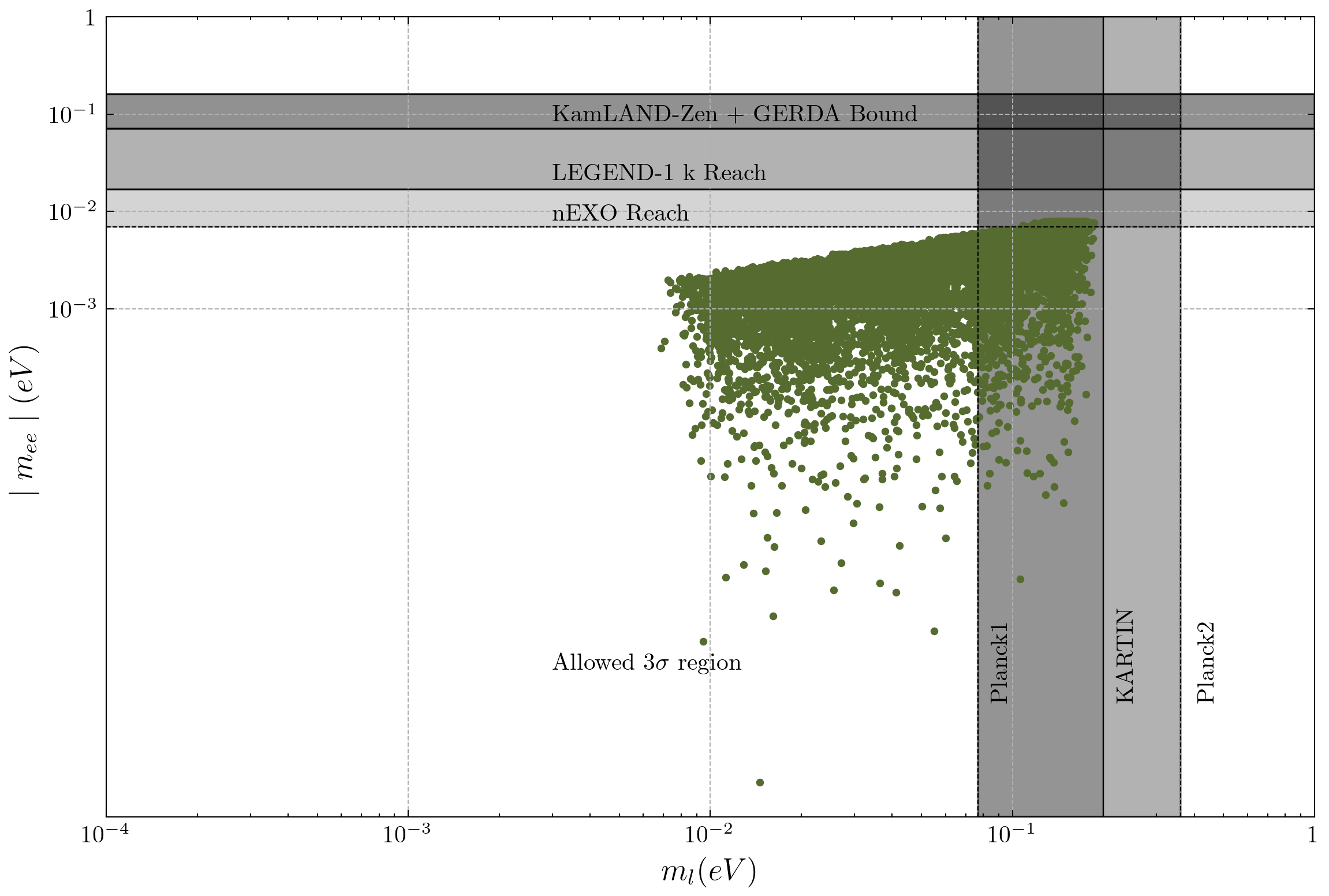}
     \end{subfigure}

  \caption{ Correlation between Effective Majorana neutrino
mass($m_{ee}$) and the lightest neutrino mass($m_{l}$).}
 \label{fig:5}
\end{figure}

Upon investigating the constrained parameter domain, we have computed the value of $\lvert m_{ee} \rvert$ in the NH scenario. Figure \ref{fig:5} illustrates variations in $\lvert m_{ee} \rvert$ corresponding to the lightest neutrino mass ($m_l$). Furthermore, it demonstrates the sensitivity range of experiments such as GERDA, KamLAND-Zen, nEXO and LEGEND-1k for neutrinoless double beta decay within the same diagram. The combined constraints from KamLAND-Zen and GERDA experiments puts an upper limit on $\lvert m_{ee} \rvert$ in the range 0.071–0.161 eV. The LEGEND-1k experiment puts a
band in $ \lvert m_{ee} \rvert$ with lowest value as 0.017 eV. The future sentivity of KATRIN $m_{lightest}$ is around $0.2 eV$ 
 \cite{katrin2001katrin}. According to the results, $\lvert m_{ee} \rvert$ falls well within the detection capabilities of these NDBD experiments in the case of the normal hierarchy.

\section{Conclusion}
\label{conc}

 Our model exploits the unique possibility of multiple allowed  yet
qualitatively different, contractions of fields charged under the $\Delta(54)$ discrete symmetry. In order to produce a neutrino mass matrix, we have proposed the $\Delta(54)$ flavor with SM Higgs boson and $Z_2 \otimes Z_3 \otimes Z_4$ symmetry. Using the Triple ISS mechanism, we developed a flavor-symmetric approach to achieve neutrino masses and mixing that align with current neutrino oscillation data. This includes accounting for the non-zero reactor angle ($\theta_{13}$) and CP violation ($\delta_{CP}$). We have incorporated additional flavons to achieve the intended mixing pattern in our study. The examined model clearly departs from the Tribimaximal (TBM) mixing pattern in the neutrino mixing matrix. The anticipated values for neutrino oscillation parameters derived from the resultant mass matrix align with the best-fit values obtained through $\chi^2$ analysis. However, the anticipated mixing angles, mass-squared differences, and the CP violation phase in the Inverted Hierarchy (IH) scenario do not concur with experimental data. In the Normal Hierarchy (NH) scenario, the model projected mixing angles indicate a preference for the upper octant of atmospheric angle $(\theta_{23})$ within the specified parameter space. \par
 
Within the framework of our $\Delta(54)$ flavor model, we conducted an in-depth examination of the Jarlskog invariant parameter ($J$) and the Neutrinoless Double Beta Decay (NDBD) phenomenon. Notably, our calculations indicate that the effective Majorana neutrino mass ($|m_{ee}|$) falls within the sensitivity range of current $0\nu \beta \beta$ experiments. Our findings are consistent with established neutrino oscillation parameters. This study lays the groundwork for future investigations, including explorations of Leptogenesis and Asymmetric Dark Matter within the context of our model

\section*{Acknowledgements} \par

HB acknowledges Tezpur University, India for Institutional Research Fellowship and to Shawan Kumar Jha for his assistance. 

\section*{Appendix}

Tensor products of $\Delta(54)$

\begin{align*}
    1_{1}\otimes S_{i} = S_{i}, \qquad  1_{2}\otimes 1_{2} = 1_{1}, \qquad 1_{2}\otimes 3_{1(1)} = 3_{2(1)} \\
    1_{2}\otimes 3_{1(2)} = 3_{2(2)} , \qquad  1_{2}\otimes 3_{2(1)} = 3_{1(1)} , \qquad 1_{2}\otimes 3_{2(2)} = 3_{1(2)} 
\end{align*}

 \begin{align*}
    \begin{pmatrix}
    a_1\\ a_2\\a_3\end{pmatrix}_{3_{1(1)}} \otimes   \begin{pmatrix}
    b_1\\b_2\\b_3\end{pmatrix}_{3_{1(1)}} = \begin{pmatrix}
    a_{1}b_{1}\\a_{2}b_{2}\\a_{3}b_{3}\end{pmatrix}_{3_{1(2)}}  \oplus \begin{pmatrix}
    a_{2}b_{3} + a_{3}b_{2} \\a_{3}b_{1} + a_{1}b_{3} \\a_{1}b_{2} + a_{2}b_{1} \end{pmatrix}_{3_{1(2)}} \oplus  \begin{pmatrix}
    a_{2}b_{3} - a_{3}b_{2} \\a_{3}b_{1} - a_{1}b_{3} \\a_{1}b_{2} - a_{2}b_{1}\end{pmatrix}_{3_{2(2)}}   
\end{align*}

  \begin{align*}
    \begin{pmatrix}
    a_1\\ a_2\\a_3\end{pmatrix}_{3_{1(2)}} \otimes   \begin{pmatrix}
    b_1\\b_2\\b_3\end{pmatrix}_{3_{1(2)}} = \begin{pmatrix}
    a_{1}b_{1}\\a_{2}b_{2}\\a_{3}b_{3}\end{pmatrix}_{3_{1(1)}}  \oplus \begin{pmatrix}
    a_{2}b_{3} + a_{3}b_{2} \\a_{3}b_{1} + a_{1}b_{3} \\a_{1}b_{2} + a_{2}b_{1} \end{pmatrix}_{3_{1(1)}} \oplus  \begin{pmatrix}
    a_{2}b_{3} - a_{3}b_{2} \\a_{3}b_{1} - a_{1}b_{3} \\a_{1}b_{2} - a_{2}b_{1}\end{pmatrix}_{3_{2(1)}}   
\end{align*}

 \begin{align*}
    \begin{pmatrix}
    a_1\\ a_2\\a_3\end{pmatrix}_{3_{2(1)}} \otimes   \begin{pmatrix}
    b_1\\b_2\\b_3\end{pmatrix}_{3_{2(1)}} = \begin{pmatrix}
    a_{1}b_{1}\\a_{2}b_{2}\\a_{3}b_{3}\end{pmatrix}_{3_{1(2)}}  \oplus \begin{pmatrix}
    a_{2}b_{3} + a_{3}b_{2} \\a_{3}b_{1} + a_{1}b_{3} \\a_{1}b_{2} + a_{2}b_{1} \end{pmatrix}_{3_{1(2)}} \oplus  \begin{pmatrix}
    a_{2}b_{3} - a_{3}b_{2} \\a_{3}b_{1} - a_{1}b_{3} \\a_{1}b_{2} - a_{2}b_{1}\end{pmatrix}_{3_{2(2)}}   
\end{align*}

 \begin{align*}
    \begin{pmatrix}
    a_1\\ a_2\\a_3\end{pmatrix}_{3_{2(2)}} \otimes   \begin{pmatrix}
    b_1\\b_2\\b_3\end{pmatrix}_{3_{2(2)}} = \begin{pmatrix}
    a_{1}b_{1}\\a_{2}b_{2}\\a_{3}b_{3}\end{pmatrix}_{3_{1(1)}}  \oplus \begin{pmatrix}
    a_{2}b_{3} + a_{3}b_{2} \\a_{3}b_{1} + a_{1}b_{3} \\a_{1}b_{2} + a_{2}b_{1} \end{pmatrix}_{3_{1(1)}} \oplus  \begin{pmatrix}
    a_{2}b_{3} - a_{3}b_{2} \\a_{3}b_{1} - a_{1}b_{3} \\a_{1}b_{2} - a_{2}b_{1}\end{pmatrix}_{3_{2(1)}} 
\end{align*}

 \begin{align*}
     \begin{pmatrix}
    a_1\\ a_2\\a_3\end{pmatrix}_{3_{1(1)}} \otimes   \begin{pmatrix}
    b_1\\b_2\\b_3\end{pmatrix}_{3_{1(2)}} = &\begin{pmatrix}
    a_{1}b_{1} + a_{2}b_{2} + a_{3}b_{3} \end{pmatrix}_{1_{1}}  \oplus \begin{pmatrix}
    a_{1}b_{1} + \omega^2 a_{2}b_{2} + \omega a_{3}b_{3 }\\ \omega a_{1}b_{1} + \omega^2 a_{2}b_{2} +  a_{3}b_{3 }\end{pmatrix}_{2_{1}} \\&
    \oplus  \begin{pmatrix}
    a_{1}b_{2} + \omega^2 a_{2}b_{3} + \omega a_{3}b_{1 }\\\omega a_{1}b_{3} + \omega^2 a_{2}b_{1} + a_{3}b_{2}\end{pmatrix}_{2_{2}}  \oplus  \begin{pmatrix}
    a_{1}b_{3} + \omega^2 a_{2}b_{1} + \omega a_{3}b_{2}\\ \omega a_{1}b_{2} + \omega^2 a_{2}b_{3} + \omega a_{3}b_{1}\end{pmatrix}_{2_{3}} \\& \oplus \begin{pmatrix}
    a_{1}b_{3} +  a_{2}b_{1} + a_{3}b_{2}\\a_{1}b_{2} +  a_{2}b_{3} + a_{3}b_{1 }\end{pmatrix}_{2_{4}} \\
\end{align*}

\begin{align*}
    \begin{pmatrix}
    a_1\\ a_2\\a_3\end{pmatrix}_{3_{1(1)}} \otimes   \begin{pmatrix}
    b_1\\b_2\\b_3\end{pmatrix}_{3_{2(1)}} = \begin{pmatrix}
    a_{1}b_{1}\\a_{2}b_{2}\\a_{3}b_{3}\end{pmatrix}_{3_{2(2)}}  \oplus \begin{pmatrix}
    a_{3}b_{2} - a_{2}b_{3} \\a_{1}b_{3} - a_{3}b_{1} \\a_{2}b_{1} - a_{1}b_{2} \end{pmatrix}_{3_{1(2)}} \oplus  \begin{pmatrix}
    a_{3}b_{2} + a_{2}b_{3} \\a_{1}b_{3} + a_{3}b_{1} \\a_{2}b_{1} + a_{1}b_{2}\end{pmatrix}_{3_{2(2)}}   
\end{align*}

 \begin{align*}
     \begin{pmatrix}
    a_1\\ a_2\\a_3\end{pmatrix}_{3_{1(1)}} \otimes   \begin{pmatrix}
    b_1\\b_2\\b_3\end{pmatrix}_{3_{2(2)}} = &\begin{pmatrix}
    a_{1}b_{1} + a_{2}b_{2} + a_{3}b_{3} \end{pmatrix}_{1_{2}}  \oplus \begin{pmatrix}
    a_{1}b_{1} + \omega^2 a_{2}b_{2} + \omega a_{3}b_{3 }\\ -\omega a_{1}b_{1} - \omega^2 a_{2}b_{2} -  a_{3}b_{3 }\end{pmatrix}_{2_{1}} \\&
    \oplus  \begin{pmatrix}
    a_{1}b_{2} + \omega^2 a_{2}b_{3} + \omega a_{3}b_{1 }\\-\omega a_{1}b_{3} - \omega^2 a_{2}b_{1} - a_{3}b_{2}\end{pmatrix}_{2_{2}}  \oplus  \begin{pmatrix}
    a_{1}b_{3} + \omega^2 a_{2}b_{1} + \omega a_{3}b_{2}\\ -\omega a_{1}b_{2} - \omega^2 a_{2}b_{3} - a_{3}b_{1}\end{pmatrix}_{2_{3}} \\& \oplus \begin{pmatrix}
    a_{1}b_{3} +  a_{2}b_{1} + a_{3}b_{2}\\ -a_{1}b_{2} - a_{2}b_{3} -  a_{3}b_{1 }\end{pmatrix}_{2_{4}} \\
\end{align*}

 \begin{align*}
     \begin{pmatrix}
    a_1\\ a_2\\a_3\end{pmatrix}_{3_{1(2)}} \otimes   \begin{pmatrix}
    b_1\\b_2\\b_3\end{pmatrix}_{3_{2(1)}} = &\begin{pmatrix}
    a_{1}b_{1} + a_{2}b_{2} + a_{3}b_{3} \end{pmatrix}_{1_{2}}  \oplus \begin{pmatrix}
    a_{1}b_{1} + \omega^2 a_{2}b_{2} + \omega a_{3}b_{3 }\\ -\omega a_{1}b_{1} - \omega^2 a_{2}b_{2} -  a_{3}b_{3 }\end{pmatrix}_{2_{1}} \\&
    \oplus  \begin{pmatrix}
    a_{1}b_{2} + \omega^2 a_{2}b_{3} + \omega a_{3}b_{1 }\\-\omega a_{1}b_{3} - \omega^2 a_{2}b_{1} - a_{3}b_{2}\end{pmatrix}_{2_{2}}  \oplus  \begin{pmatrix}
    a_{1}b_{3} + \omega^2 a_{2}b_{1} + \omega a_{3}b_{2}\\ -\omega a_{1}b_{2} - \omega^2 a_{2}b_{3} - a_{3}b_{1}\end{pmatrix}_{2_{3}} \\& \oplus \begin{pmatrix}
    a_{1}b_{3} +  a_{2}b_{1} + a_{3}b_{2}\\ -a_{1}b_{2} - a_{2}b_{3} -  a_{3}b_{1 }\end{pmatrix}_{2_{4}} \\
\end{align*}

 \begin{align*}
    \begin{pmatrix}
    a_1\\ a_2\\a_3\end{pmatrix}_{3_{1(2)}} \otimes   \begin{pmatrix}
    b_1\\b_2\\b_3\end{pmatrix}_{3_{2(2)}} = \begin{pmatrix}
    a_{1}b_{1}\\a_{2}b_{2}\\a_{3}b_{3}\end{pmatrix}_{3_{2(1)}}  \oplus \begin{pmatrix}
    a_{3}b_{2} - a_{2}b_{3} \\a_{1}b_{3} - a_{3}b_{1} \\a_{2}b_{1} - a_{1}b_{2} \end{pmatrix}_{3_{1(1)}} \oplus  \begin{pmatrix}
    a_{3}b_{2} + a_{2}b_{3} \\a_{1}b_{3} + a_{3}b_{1} \\a_{2}b_{1} + a_{1}b_{2}\end{pmatrix}_{3_{2(1)}}   
\end{align*}

  \begin{align*}
     \begin{pmatrix}
    a_1\\ a_2\\a_3\end{pmatrix}_{3_{2(1)}} \otimes   \begin{pmatrix}
    b_1\\b_2\\b_3\end{pmatrix}_{3_{2(2)}} = &\begin{pmatrix}
    a_{1}b_{1} + a_{2}b_{2} + a_{3}b_{3} \end{pmatrix}_{1_{1}}  \oplus \begin{pmatrix}
    a_{1}b_{1} + \omega^2 a_{2}b_{2} + \omega a_{3}b_{3 }\\ \omega a_{1}b_{1} + \omega^2 a_{2}b_{2} +  a_{3}b_{3 }\end{pmatrix}_{2_{1}} \\&
    \oplus  \begin{pmatrix}
    a_{1}b_{2} + \omega^2 a_{2}b_{3} + \omega a_{3}b_{1 }\\ \omega a_{1}b_{3} + \omega^2 a_{2}b_{1} + a_{3}b_{2}\end{pmatrix}_{2_{2}}  \oplus  \begin{pmatrix}
    a_{1}b_{3} + \omega^2 a_{2}b_{1} + \omega a_{3}b_{2}\\ \omega a_{1}b_{2} + \omega^2 a_{2}b_{3} + a_{3}b_{1}\end{pmatrix}_{2_{3}} \\& \oplus \begin{pmatrix}
    a_{1}b_{3} +  a_{2}b_{1} + a_{3}b_{2}\\ a_{1}b_{2} + a_{2}b_{3} +  a_{3}b_{1 }\end{pmatrix}_{2_{4}} \\
\end{align*}

 \begin{align*}
    \begin{pmatrix}
    a_1\\ a_2 \end{pmatrix}_{2_{s}} \otimes   \begin{pmatrix}
    b_1\\ b_2 \end{pmatrix}_{2_{s}} = \begin{pmatrix}
    a_1 b_2 +  a_2 b_1 \end{pmatrix}_{1_{1}} \oplus \begin{pmatrix}
    a_1 b_2 -  a_2 b_1\end{pmatrix}_{1_{2}} \oplus \begin{pmatrix}
    a_2 b_2 \\ a_1 b_1\end{pmatrix}_{2_{s}}
\end{align*}

\begin{align*}
    \begin{pmatrix}
    a_1\\ a_2\end{pmatrix}_{2_{1}} \otimes   \begin{pmatrix}
    b_1\\b_2\end{pmatrix}_{2_{2}} = \begin{pmatrix}
    a_{2}b_{2}\\a_{1}b_{1}\end{pmatrix}_{2_{3}}  \oplus \begin{pmatrix}
     a_{2}b_{1}\\a_{1}b_{2} \end{pmatrix}_{2_{4}} 
\end{align*}

\begin{align*}
    \begin{pmatrix}
    a_1\\ a_2\end{pmatrix}_{2_{1}} \otimes   \begin{pmatrix}
    b_1\\b_2\end{pmatrix}_{2_{3}} = \begin{pmatrix}
    a_{2}b_{2}\\a_{1}b_{1}\end{pmatrix}_{2_{2}}  \oplus \begin{pmatrix}
     a_{2}b_{1}\\a_{1}b_{2} \end{pmatrix}_{2_{4}} 
\end{align*}

\begin{align*}
    \begin{pmatrix}
    a_1\\ a_2\end{pmatrix}_{2_{1}} \otimes   \begin{pmatrix}
    b_1\\b_2\end{pmatrix}_{2_{4}} = \begin{pmatrix}
    a_{1}b_{2}\\a_{2}b_{1}\end{pmatrix}_{2_{2}}  \oplus \begin{pmatrix}
     a_{1}b_{1}\\a_{2}b_{2} \end{pmatrix}_{2_{3}} 
\end{align*}

\begin{align*}
    \begin{pmatrix}
    a_1\\ a_2\end{pmatrix}_{2_{2}} \otimes   \begin{pmatrix}
    b_1\\b_2\end{pmatrix}_{2_{3}} = \begin{pmatrix}
    a_{2}b_{2}\\a_{1}b_{1}\end{pmatrix}_{2_{1}}  \oplus \begin{pmatrix}
     a_{1}b_{2}\\a_{2}b_{1} \end{pmatrix}_{2_{4}} 
\end{align*}

\begin{align*}
    \begin{pmatrix}
    a_1\\ a_2\end{pmatrix}_{2_{2}} \otimes   \begin{pmatrix}
    b_1\\b_2\end{pmatrix}_{2_{4}} = \begin{pmatrix}
    a_{1}b_{1}\\a_{2}b_{2}\end{pmatrix}_{2_{1}}  \oplus \begin{pmatrix}
     a_{1}b_{2}\\a_{2}b_{1} \end{pmatrix}_{2_{3}} 
\end{align*}

\begin{align*}
    \begin{pmatrix}
    a_1\\ a_2\end{pmatrix}_{2_{3}} \otimes   \begin{pmatrix}
    b_1\\b_2\end{pmatrix}_{2_{4}} = \begin{pmatrix}
    a_{1}b_{2}\\a_{2}b_{1}\end{pmatrix}_{2_{1}}  \oplus \begin{pmatrix}
     a_{1}b_{1}\\a_{2}b_{2} \end{pmatrix}_{2_{2}} 
\end{align*}\\

\bibliographystyle{naturemag} 
\bibliography{references}

\begin{thebibliography}{10}
\expandafter\ifx\csname url\endcsname\relax
  \def\url#1{\texttt{#1}}\fi
\expandafter\ifx\csname urlprefix\endcsname\relax\def\urlprefix{URL }\fi
\providecommand{\bibinfo}[2]{#2}
\providecommand{\eprint}[2][]{\url{#2}}

\bibitem{DayaBay:2012fng}
\bibinfo{author}{An, F.~P.} \emph{et~al.}
\newblock \bibinfo{title}{{Observation of electron-antineutrino disappearance at Daya Bay}}.
\newblock \emph{\bibinfo{journal}{Phys. Rev. Lett.}} \textbf{\bibinfo{volume}{108}}, \bibinfo{pages}{171803} (\bibinfo{year}{2012}).
\newblock \eprint{1203.1669}.

\bibitem{RENO:2012mkc}
\bibinfo{author}{Ahn, J.~K.} \emph{et~al.}
\newblock \bibinfo{title}{{Observation of Reactor Electron Antineutrino Disappearance in the RENO Experiment}}.
\newblock \emph{\bibinfo{journal}{Phys. Rev. Lett.}} \textbf{\bibinfo{volume}{108}}, \bibinfo{pages}{191802} (\bibinfo{year}{2012}).
\newblock \eprint{1204.0626}.

\bibitem{MINOS:2011amj}
\bibinfo{author}{Adamson, P.} \emph{et~al.}
\newblock \bibinfo{title}{{Improved search for muon-neutrino to electron-neutrino oscillations in MINOS}}.
\newblock \emph{\bibinfo{journal}{Phys. Rev. Lett.}} \textbf{\bibinfo{volume}{107}}, \bibinfo{pages}{181802} (\bibinfo{year}{2011}).
\newblock \eprint{1108.0015}.

\bibitem{DoubleChooz:2011ymz}
\bibinfo{author}{Abe, Y.} \emph{et~al.}
\newblock \bibinfo{title}{{Indication of Reactor $\bar{\nu}_e$ Disappearance in the Double Chooz Experiment}}.
\newblock \emph{\bibinfo{journal}{Phys. Rev. Lett.}} \textbf{\bibinfo{volume}{108}}, \bibinfo{pages}{131801} (\bibinfo{year}{2012}).
\newblock \eprint{1112.6353}.

\bibitem{T2K:2011ypd}
\bibinfo{author}{Abe, K.} \emph{et~al.}
\newblock \bibinfo{title}{{Indication of Electron Neutrino Appearance from an Accelerator-produced Off-axis Muon Neutrino Beam}}.
\newblock \emph{\bibinfo{journal}{Phys. Rev. Lett.}} \textbf{\bibinfo{volume}{107}}, \bibinfo{pages}{041801} (\bibinfo{year}{2011}).
\newblock \eprint{1106.2822}.

\bibitem{Furry:1939qr}
\bibinfo{author}{Furry, W.~H.}
\newblock \bibinfo{title}{{On transition probabilities in double beta-disintegration}}.
\newblock \emph{\bibinfo{journal}{Phys. Rev.}} \textbf{\bibinfo{volume}{56}}, \bibinfo{pages}{1184--1193} (\bibinfo{year}{1939}).

\bibitem{DellOro:2016tmg}
\bibinfo{author}{Dell'Oro, S.}, \bibinfo{author}{Marcocci, S.}, \bibinfo{author}{Viel, M.} \& \bibinfo{author}{Vissani, F.}
\newblock \bibinfo{title}{{Neutrinoless double beta decay: 2015 review}}.
\newblock \emph{\bibinfo{journal}{Adv. High Energy Phys.}} \textbf{\bibinfo{volume}{2016}}, \bibinfo{pages}{2162659} (\bibinfo{year}{2016}).
\newblock \eprint{1601.07512}.

\bibitem{barman2023non}
\bibinfo{author}{Barman, A.}, \bibinfo{author}{Francis, N.~K.}, \bibinfo{author}{Thapa, B.} \& \bibinfo{author}{Nath, A.}
\newblock \bibinfo{title}{{Non-zero $\theta_{13}$, CP-violation and Neutrinoless Double Beta Decay for Neutrino Mixing in the $A_{4}\times Z_{2} \times Z_{3}$ Flavor Symmetry Model}}.
\newblock \emph{\bibinfo{journal}{International Journal of Modern Physics A}}  (\bibinfo{year}{2023}).

\bibitem{barman2023neutrino}
\bibinfo{author}{Barman, A.}, \bibinfo{author}{Francis, N.~K.} \& \bibinfo{author}{Bora, H.}
\newblock \bibinfo{title}{{Neutrino Mixing Phenomenology: $A_4$ Discrete Flavor Symmetry with Type-I Seesaw Mechanism}}.
\newblock \emph{\bibinfo{journal}{arXiv preprint arXiv:2306.11461}}  (\bibinfo{year}{2023}).

\bibitem{Ma:2006km}
\bibinfo{author}{Ma, E.}
\newblock \bibinfo{title}{{Verifiable radiative seesaw mechanism of neutrino mass and dark matter}}.
\newblock \emph{\bibinfo{journal}{Phys. Rev. D}} \textbf{\bibinfo{volume}{73}}, \bibinfo{pages}{077301} (\bibinfo{year}{2006}).
\newblock \eprint{hep-ph/0601225}.

\bibitem{Vien:2014pta}
\bibinfo{author}{Vien, V.~V.} \& \bibinfo{author}{Long, H.~N.}
\newblock \bibinfo{title}{{Neutrino mixing with nonzero $\theta_{13}$ and CP violation in the 3-3-1 model based on $A_4$ flavor symmetry}}.
\newblock \emph{\bibinfo{journal}{Int. J. Mod. Phys. A}} \textbf{\bibinfo{volume}{30}}, \bibinfo{pages}{1550117} (\bibinfo{year}{2015}).
\newblock \eprint{1405.4665}.

\bibitem{thapa2021resonant}
\bibinfo{author}{Thapa, B.} \& \bibinfo{author}{Francis, N.~K.}
\newblock \bibinfo{title}{{Resonant leptogenesis and \uppercase{TM}$_1$ mixing in minimal type-\uppercase{I} seesaw model with \uppercase{S}$_4$ symmetry}}.
\newblock \emph{\bibinfo{journal}{The European Physical Journal C}} \textbf{\bibinfo{volume}{81}}, \bibinfo{pages}{1--8} (\bibinfo{year}{2021}).

\bibitem{Ma:2007wu}
\bibinfo{author}{Ma, E.}
\newblock \bibinfo{title}{{Near tribimaximal neutrino mixing with $\Delta(27)$ symmetry}}.
\newblock \emph{\bibinfo{journal}{Phys. Lett. B}} \textbf{\bibinfo{volume}{660}}, \bibinfo{pages}{505--507} (\bibinfo{year}{2008}).
\newblock \eprint{0709.0507}.

\bibitem{de2007neutrino}
\bibinfo{author}{de~Medeiros~Varzielas, I.}, \bibinfo{author}{King, S.} \& \bibinfo{author}{Ross, G.}
\newblock \bibinfo{title}{Neutrino tri-bi-maximal mixing from a non-abelian discrete family symmetry}.
\newblock \emph{\bibinfo{journal}{Physics Letters B}} \textbf{\bibinfo{volume}{648}}, \bibinfo{pages}{201--206} (\bibinfo{year}{2007}).

\bibitem{Harrison:2014jqa}
\bibinfo{author}{Harrison, P.~F.}, \bibinfo{author}{Krishnan, R.} \& \bibinfo{author}{Scott, W.~G.}
\newblock \bibinfo{title}{{Deviations from tribimaximal neutrino mixing using a model with $\Delta(27)$ symmetry}}.
\newblock \emph{\bibinfo{journal}{Int. J. Mod. Phys. A}} \textbf{\bibinfo{volume}{29}}, \bibinfo{pages}{1450095} (\bibinfo{year}{2014}).
\newblock \eprint{1406.2025}.

\bibitem{CarcamoHernandez:2016piw}
\bibinfo{author}{C\'arcamo~Hern\'andez, A.~E.}, \bibinfo{author}{Long, H.~N.} \& \bibinfo{author}{Vien, V.~V.}
\newblock \bibinfo{title}{{A 3-3-1 model with right-handed neutrinos based on the $\Delta (27) $ family symmetry}}.
\newblock \emph{\bibinfo{journal}{Eur. Phys. J. C}} \textbf{\bibinfo{volume}{76}}, \bibinfo{pages}{242} (\bibinfo{year}{2016}).
\newblock \eprint{1601.05062}.

\bibitem{loualidi2021trimaximal}
\bibinfo{author}{Loualidi, M.}
\newblock \bibinfo{title}{{Trimaximal mixing with one texture zero from type II seesaw and $\Delta(54)$ family symmetry}}.
\newblock \emph{\bibinfo{journal}{arXiv preprint arXiv:2104.13734}}  (\bibinfo{year}{2021}).

\bibitem{ishimori2009lepton}
\bibinfo{author}{Ishimori, H.}, \bibinfo{author}{Kobayashi, T.}, \bibinfo{author}{Okada, H.}, \bibinfo{author}{Shimizu, Y.} \& \bibinfo{author}{Tanimoto, M.}
\newblock \bibinfo{title}{{Lepton flavor model from $\Delta$ (54) symmetry}}.
\newblock \emph{\bibinfo{journal}{Journal of High Energy Physics}} \textbf{\bibinfo{volume}{2009}}, \bibinfo{pages}{011} (\bibinfo{year}{2009}).

\bibitem{vien2021extension}
\bibinfo{author}{Vien, V.~V.}
\newblock \bibinfo{title}{An extension of the standard model with symmetry for quark masses and mixings}.
\newblock \emph{\bibinfo{journal}{Physics of Atomic Nuclei}} \textbf{\bibinfo{volume}{84}}, \bibinfo{pages}{179--183} (\bibinfo{year}{2021}).

\bibitem{nilles2018cp}
\bibinfo{author}{Nilles, H.~P.}, \bibinfo{author}{Ratz, M.}, \bibinfo{author}{Trautner, A.} \& \bibinfo{author}{Vaudrevange, P.~K.}
\newblock \bibinfo{title}{Cp violation from string theory}.
\newblock \emph{\bibinfo{journal}{Physics Letters B}} \textbf{\bibinfo{volume}{786}}, \bibinfo{pages}{283--287} (\bibinfo{year}{2018}).

\bibitem{bora2023neutrino}
\bibinfo{author}{Bora, H.}, \bibinfo{author}{Francis, N.~K.}, \bibinfo{author}{Barman, A.} \& \bibinfo{author}{Thapa, B.}
\newblock \bibinfo{title}{{Neutrino Mass Model in the Context of $\Delta(54) \otimes Z_2 \otimes Z_3 \otimes Z_4$ Flavor Symmetries with Inverse Seesaw Mechanism}}.
\newblock \emph{\bibinfo{journal}{Physics Letters B}} \bibinfo{pages}{138329} (\bibinfo{year}{2023}).

\bibitem{2010}
\bibinfo{author}{Altarelli, G.} \& \bibinfo{author}{Feruglio, F.}
\newblock \bibinfo{title}{Discrete flavor symmetries and models of neutrino mixing}.
\newblock \emph{\bibinfo{journal}{Reviews of Modern Physics}} \textbf{\bibinfo{volume}{82}}, \bibinfo{pages}{2701–2729} (\bibinfo{year}{2010}).
\newblock \urlprefix\url{http://dx.doi.org/10.1103/RevModPhys.82.2701}.

\bibitem{chulia2021inverse}
\bibinfo{author}{Chuli{\'a}, S.~C.}, \bibinfo{author}{Srivastava, R.} \& \bibinfo{author}{Vicente, A.}
\newblock \bibinfo{title}{The inverse seesaw family: Dirac and majorana}.
\newblock \emph{\bibinfo{journal}{Journal of High Energy Physics}} \textbf{\bibinfo{volume}{2021}}, \bibinfo{pages}{1--29} (\bibinfo{year}{2021}).

\bibitem{estebannufit}
\bibinfo{author}{Esteban, I.} \emph{et~al.}
\newblock \bibinfo{title}{Nufit 6.0: Three-neutrino fit based on data available in september 2024}  (\bibinfo{year}{2024}).

\bibitem{lei2020minimally}
\bibinfo{author}{Lei, M.} \& \bibinfo{author}{Wells, J.~D.}
\newblock \bibinfo{title}{{Minimally modified $A_4$ Altarelli-Feruglio model for neutrino masses and mixings and its experimental consequences}}.
\newblock \emph{\bibinfo{journal}{Physical Review D}} \textbf{\bibinfo{volume}{102}}, \bibinfo{pages}{016023} (\bibinfo{year}{2020}).

\bibitem{katrin2001katrin}
\bibinfo{author}{Collaboration, K.} \emph{et~al.}
\newblock \bibinfo{title}{Katrin: A next generation tritium beta decay experiment with sub-ev sensitivity for the electron neutrino mass}.
\newblock \emph{\bibinfo{journal}{arXiv preprint hep-ex/0109033}}  (\bibinfo{year}{2001}).

\end{thebibliography}

\end{document}